\pdfoutput=1 %force arxiv to use pdflatex
\documentclass[10pt, a4paper]{article}
\usepackage{lrec2022} % this is the new LREC2022 Style
\usepackage{multibib}
\newcites{languageresource}{Language Resources}
\usepackage{graphicx}
\usepackage{tabularx}
\usepackage{soul}
\usepackage{amsmath}
\usepackage{booktabs}
\usepackage{multirow}
\usepackage{todonotes} 
\usepackage{pgfplots, pgfplotstable}
\pgfplotsset{compat=newest}
\usepgfplotslibrary{groupplots} 
\newcommand{\bftab}[1]{{\fontseries{b}\selectfont #1}} %command I use do boldface numbers in table so they do not get extended

\usepackage{titlesec}
\titleformat{\section}{\normalfont\large\bfseries\center}{\thesection.}{1em}{}
\titleformat{\subsection}{\normalfont\SmallTitleFont\bfseries\raggedright}{\thesubsection.}{1em}{}
\titleformat{\subsubsection}{\normalfont\normalsize\bfseries\raggedright}{\thesubsubsection.}{1em}{}
\renewcommand\thesection{\arabic{section}}
\renewcommand\thesubsection{\thesection.\arabic{subsection}}
\renewcommand\thesubsubsection{\thesubsection.\arabic{subsubsection}}
\usepackage{epstopdf}
\usepackage[utf8]{inputenc}

\usepackage{hyperref}
\usepackage{xstring}

\usepackage{color}

\title{Human and Automatic Speech Recognition Performance\\on German Oral History Interviews}
 
\name{\shortstack{Michael Gref$^{~1}$, Nike Matthiesen$^{2}$, Christoph Schmidt$^{1}$, Sven Behnke$^{1,3}$, Joachim Köhler$^{1}$}} 
\address{$^1$Fraunhofer Institute for Intelligent Analysis and Information Systems (IAIS), Germany\\
	$^2$Haus der Geschichte der Bundesrepublik Deutschland Foundation (HdG), Bonn, Germany\\
	$^3$Autonomous Intelligent Systems (AIS), Computer Science Institute VI, Univ. of Bonn, Germany\\
	\{michael.gref, christoph.andreas.schmidt, sven.behnke, joachim.koehler\}@iais.fraunhofer.de\\
	  matthiesen@hdg.de}

\abstract{
Automatic speech recognition systems have accomplished remarkable improvements in transcription accuracy in recent years. On some domains, models now achieve near-human performance. However, transcription performance on oral history has not yet reached human accuracy. In the present work, we investigate how large this gap between human and machine transcription still is. For this purpose, we analyze and compare transcriptions of three humans on a new oral history data set. We estimate a human word error rate of 8.7\,\% for recent German oral history interviews with clean acoustic conditions. For comparison with recent machine transcription accuracy, we present experiments on the adaptation of an acoustic model achieving near-human performance on broadcast speech. We investigate the influence of different adaptation data on robustness and generalization for clean and noisy oral history interviews. We optimize our acoustic models by 5 to 8\,\% relative for this task and achieve 23.9\,\% WER on noisy and 15.6\,\% word error rate on clean oral history interviews. 
 \\ \newline  }

\begin{document}

\maketitleabstract

\section{Introduction}

Automatic speech recognition (ASR) has achieved remarkable improvements in transcription accuracy in recent years, enabling diverse new applications. For certain domains, systems have achieved or surpassed a human-near transcription accuracy, e.g., on conversational speech of the English Switchboard ASR task \cite{Saon.2017.IBM_SWBD,Xiong.2017.HumanPartiyInASR}. 

One application that seemed unthinkable half a decade ago is the high-quality transcription of \textit{oral history interviews}. In the humanities, oral history refers to conducting and analyzing interviews with contemporary witnesses to historical events. Oral history archives are often large audiovisual data repositories composing numerous interviews, often a few hours in length per interview. Recent ASR systems simplified assessing these archives by generating time-aligned transcriptions automatically. A few years ago, the recognition performance of ASR systems for these interviews was still in a range that, at best, allowed for a keyword-based search. However, today's adapted systems achieve a recognition quality for many interviews where the transcript, while not entirely error-free, is reasonably readable for indexing the content in search queries, searching for quotes in interviews, or utilizing the transcript in subsequent natural language processing (NLP) analysis steps. With manual correction, these ASR transcripts can also be used, for example, as subtitles to the oral history video on digital platforms. For instance, in a recent survey, \newcite{Pessanha.2022.ComputOralHistoryOverview} identify automatic speech recognition as one of the possible key technologies, in combination with NLP systems or approaches for analyzing non-verbal cues, that can improve the future access of archives. In particular, these approaches can offer insights that exceed approaches based on human work alone.

Despite this progress in ASR, oral history interviews are still a big challenge for automatic transcription systems due to their heterogeneity in terms of language, audio, speech type, dialects, and speakers. Therefore, this paper deals with the question of where we stand today with respect to the recognition performance of ASR systems. In particular, we are concerned with how automatic transcription relates to human transcription---both of which are not error-free. We consider this question crucial for assessing future work in this area, particularly the word error rate with which to achieve or exceed human performance. 

We investigate which mistakes humans make when transcribing or, more precisely, correcting ASR transcriptions. Additionally, we estimate a human word error rate using various human corrections of raw ASR transcripts. This human word error rate measures differences and ambiguities in different human transcriptions of the same interviews. We are investigating these questions for a new German oral history data set of videos made publicly available in an online portal. Furthermore, this paper presents our most recent ASR models adapted to oral history and discusses the influence of different properties of the interviews on training performance.

\section{Related Work}

Best to our knowledge, no work estimated a human word error rate for oral history interviews so far---particularly for German Oral History data. However, such a measure has been estimated for individual data sets. \newcite{Lippmann.1997.HumanWER} performed one of the early works comparing ASR with human word error rates. The authors estimate the human word error rate for different domains using common English corpora, such as the WSJ for read-speech and Switchboard for conversational speech. The authors report a human error rate of about 1\,\% for the Wallstreet Journal Corpus (WSJ) \citelanguageresource{Paul.1992.WSJCorpus} and 4\,\% for Switchboard \citelanguageresource{Godfrey.1992.Switchboard}.

With the enormously increased recognition performance of ASR in the last decade, \newcite{Xiong.2017.HumanPartiyInASR} and \newcite{Saon.2017.IBM_SWBD} reconsidered the human word error rate on the English Switchboard corpus. The reported human error rate of these works was in the range of 5.1---5.9\,\%. While proposed as an error rate for this particular corpus, this human word error rate was sometimes misconstrued as a general human word error rate in the general public. A detailed overview of Xiong et al.'s and Saon et al.'s approaches for human word error rate estimation is given in Section \ref{sec:human_wer}, where we present and compare our approach for the estimation.

In the following, we give an overview of related work that applies and examines ASR for oral history interviews in different languages. This aims to put word error rates on this challenging domain achieved by different ASR systems over time into perspective.

The research on ASR for oral history interviews started in 2002 with the MALACH project \cite{Psutka.2002.MalachProject}. Subsequent studies investigated improvements and error sources of ASR systems from the respective time on the MALACH data \cite{Ramabhadrana.2003.MalachEarlyWork1}. \newcite{Siohan.2004.MalachErrorAnalysis} investigated possible speech recognition error sources for this data.

Over the years, ASR for oral history has been studied in various languages. This was done partly on public data from the MALACH corpus and partly on proprietary in-house data. Even in recent times, a comparatively high word error rate (WER) on the oral history domain, compared to other ASR tasks, characterizes most works. \newcite{Hessen.2013.EinsatzASROralHistory} describe the use of ASR to transcribe Dutch oral history archives. The authors state the WER is above 40\,\% for Dutch oral history interviews at the time of publishing. 

\newcite{Salesky.2016.OralHistoryKeywordSearch} studied \textit{keyword ASR} for English oral history data collected by \textit{StoryCorps}, an American non-profit organization. The ASR results are used, among others, to assess and compare a \textit{human search capability} on oral history data. The authors perform experiments with 100 hours of \textit{in-domain} training data and with \textit{out-of-domain} training data. They achieved a 38.5\,\% WER on the 50-hour evaluation set with in-domain training data. Experiments with only out-of-domain training data yield between 49.5\,\% and 68.1\,\% WER. 

\newcite{Zajic.2018.ProessingOralHistory} studied speech recognition with Czech interviews from MALACH. The authors trained a hybrid acoustic model with 84 hours of Czech MALACH recordings. For evaluation, ten interviews with overall 60,000 running words were used. Using a language model (LM) that combines texts from MALACH with additional text resources, the authors achieved 42.0\,\% WER on their test data. In an \textit{oracle experiment}, \newcite{Zajic.2018.ProessingOralHistory} used the transcripts of the test data only for language model training and achieved 19.5\,\% WER. The authors state that this oracle experiment shows the current performance upper bound of the ASR system for the author's data.

\newcite{Zajic.2018.ProessingOralHistory} state, often in line with the aforementioned works, that it is challenging to design an ASR system for oral history interviews that is accurate enough due to the nature of the interviews. The interviewees are usually elderly people, their spontaneous speech is frequently accented, and they are often emotional due to the nature of their experiences. Also, the oral history interviews' speech quality is relatively low with many disfluencies and non-speech events such as crying and laughing. The regular use of colloquial words also negatively impacts speech recognition. 

The work on the English MALACH corpus continued with \newcite{Pincheny.2019.MalachASRCorpus}, who propose data from the MALACH corpus as a new ASR challenge for English oral history. The authors report reference results of a 25.9\,\% WER (with a conventional LM) and 21.7\,\% (with an LSTM-LM) using sequence-discriminative sMBR training for hybrid acoustic models. For training, the authors used 176 hours of manually annotated oral history interviews. The authors are in line with the aforementioned works that the challenges of oral history interviews are still open problems for modern ASR---even in 2019. 

We examined German oral history interviews using the KA3 data set in our prior works. In our initial works, we focused on the acoustic challenges of the recordings \cite{Gref.2018.oralhistory02}. Room reverberation posed a particular challenge for our data. In more recent works, we have focused on improving spontaneous speech and interviewees' speech characteristics by adapting the acoustic model. In our latest paper, we achieved 25.2\,\% WER on the KA3 data set with cross-lingual adaptation on 25 hours of forced-aligned oral history interviews \cite{Gref.2020.MultiStageCrossLingual}.

\section{The \textit{HdG} Oral History Data Set}
\label{sect:hdg_dataset}

\textit{Zeitzeugenportal}\footnote{\url{https://www.zeitzeugen-portal.de}} is a German online service by \textit{Haus der Geschichte} (House of the History) Foundation (HdG) and includes a central collection of contemporary German oral history interviews.  More than 8,000 clips from around 1,000 interviews can already be found at \textit{Zeitzeugenportal}. The large database does not only offer interesting historical content, but also allows the viewer to empathize with what he or she has experienced through emotionally charged stories. This online service gives users the chance to hear about people’s stories at any time. In the following section, we present the research project as well as the data set for the annotations.  

We selected 10 hours of German oral history interviews from the HdG \textit{Zeitzeugenportal} for our experiments. Our \textit{HdG data set} comprises 164 different interview videos of 147 distinct interviewees. Thus, on average, the interviews in our data set have a length of 3.6 minutes.

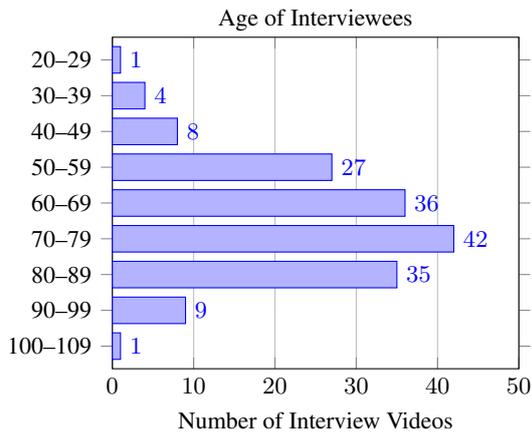
\begin{figure}[t]
	\begin{center}\vspace{-1mm}
		\begin{tikzpicture}
			\begin{axis}[
				title={Age of Interviewees},
				symbolic y coords={100--109,90--99,80--89,70--79,60--69,50--59,40--49,30--39,20--29},
%				height=0.3\linewidth, 
				xlabel={Number of Interview Videos},
				title style={yshift=-2.5mm},
				style={font=\footnotesize},
				xbar, 
				xmin=0, 
				xmax=50,
				xmajorgrids,
				width=0.9\linewidth,   
				ytick distance=1,
				enlarge y limits={abs=0.3cm},
				nodes near coords, nodes near coords align={horizontal},
				bar width=10pt,
				]
				\addplot coordinates {
					(1,20--29)
					(4,30--39)
					(8,40--49)
					(27,50--59)
					(36,60--69)
					(42,70--79)
					(35,80--89)
					(9,90--99)
					(1,100--109)
				};
			\end{axis}
		\end{tikzpicture} \vspace{-2.5mm}
		\caption{Age distribution in the HdG interview dataset. No age information was available for one of the 164 videos.}
		\label{fig:hdgset_age_dist}
	\end{center}
\end{figure}

The selected interviews were recorded between 2010 and 2020. Thus, the selection is representative of the more recent videos on the portal. It includes 66 interviews with professional speakers, who pursue a representative profession, and 98 interviews with non-professional speakers.

In addition, we aimed to represent different emotions in the selection of videos and create a heterogeneous data set in terms of age and gender. The age distribution of interviewees in the data set is shown in Figure \ref{fig:hdgset_age_dist}. We have a strong focus on interviewees between the age of 50 to 89 years as these are the most frequent interviews in the archive. Nevertheless, we have deliberately included videos of younger and older interviewees.

Throughout the entire archive, male interviewees make up most of the data. However, a representative selection would underrepresent female interviewees. Therefore, we have included additional female speakers in the HdG data set. Overall, the HdG data set contains videos with 104 male and 60 female interviewees.

After the annotation and transcription, the HdG data set is split into speaker-independent training, development, and test subset for model training and evaluation. An overview of these sets is given in Table \ref{tab:data_set_overview}. Overall, 358 segments with roughly 0.5 hours could not be annotated and were removed from the data set. These were, for example, segments containing only intros, fade-ins, or pauses.

\begin{table}[t]
	\centering
	\begin{tabular}{lrrr}
		\toprule
		\textbf{Set}		& \textbf{Videos} 	&  \textbf{Segments} &   \textbf{Hours}  \\ 
		\midrule	
		HdG Training		& 104	& 	1,863	& 	6.35 \\
		HdG Development		& 27	& 	430		& 	1.44 \\
		HdG Test		& 33	& 	471		& 	1.74 \\
		\midrule
		KA3 Test		& 35	& 	2,392	&	3.52 \\ 
		\bottomrule
	\end{tabular}
	\caption{Overview of HdG oral history data sets after annotation and split into speaker-independent subsets. The KA3 data set is included for comparison.}
	\label{tab:data_set_overview}
\end{table} 

The HdG portal's interviews are somewhat different from the interviews in the \textit{KA3 data set} \citelanguageresource{Gref.2018.oralhistory01} we utilized in our prior works in the \textit{KA3 project} \cite{Leh.2018.viewjournal_KA3}. The KA3 oral history interviews represent the archive \textit{Deutsches Gedächtnis} (German Memory), University of Hagen (Germany). While the archive works with full-length interviews with an average length of 3.5 hours, the \textit{Zeitzeugenportal} shows thematic clips of interviews with a length of 3 to 5 minutes. The interviews in this \textit{KA3 data set} were recorded between 1980 and 2012 and represent the archive's wide range of interviews with respect to recording technology. Since the interviews in our HdG data set are recorded in more recent times, often with more recent or professional equipment, the HdG data set overall have better audio recording quality than the KA3 data.

In other aspects, such as language style, age of the interviewees, dialects, and topics, the HdG and KA3 data are quite similar. Thus, we assume that the conclusions of the experiments can be applied to other data as well. The KA3 data set was transcribed entirely by hand and is used in our prior works to evaluate the performance of ASR systems. It does not include any annotation of emotion or sentiment. To obtain a robust statement about the real-world performance of our ASR systems on oral history data, we evaluate our systems on both the KA3 and HdG data sets later in this work. The difference in performance on the data sets may provide insight into the impact of audio quality and video age of oral history interviews on ASR performance.

The data set is not published and is only used in-house due to the General Data Protection Regulation and the personal rights of the interviewees. 

\section{Human Word Error Rate Estimation}
\label{sec:human_wer}

In the following section, we describe our approach to computing the human word error rate, the pipeline for our study, the results, and conclude with a discussion of transcription errors.

\subsection{Annotation Approach}
The performance of ASR systems is usually measured in terms of the \textit{word error rate} (\textit{WER}), which is the Levenshtein (or edit) distance of words between the reference transcription (Ref) and the ASR system's hypothesis (Hyp). Since this reference is usually created by humans, it may be subject to some limitations. It is not only necessary to take into account that humans can make mistakes when transcribing. Unlike some annotations in other machine learning domains, the transcription of spontaneous speech has a high degree of inherent ambiguity, cf.\ \newcite{Stolcke.2017.MicrosoftComparingHumanAndASR}. 

In our prior work, we observed that humans often tend to transcribe what they understand and not necessarily what was actually said. It seems that speech errors in spontaneous speech are often unconsciously overheard and corrected. Speech recognition systems are usually more precise in this respect and transcribe mispronounced words, repeated words, and slips of the tongue precisely as they were uttered. It depends on the further usage of the ASR transcript, whether such a phonetically exact transcription or more human-like, corrected transcription is to be considered correct. This also applies, for example, to the transcription of hesitations. For linguistics and specific historical research questions, a transcription as exact as possible of what was actually said is relevant. A human-like transcription is more desirable for other applications, e.g., for subtitling videos or further processing by NLP systems. We are currently striving for phonetically exact transcription for our current task, including hesitations. Correction of the transcript is performed in subsequent post-processing steps, which we do not consider part of the ASR.

These issues are reflected, among others, when comparing the differences between human transcribers on the same speech data. Human transcription was studied, for instance, by \newcite{Xiong.2017.HumanPartiyInASR} and \newcite{Saon.2017.IBM_SWBD} for the English Switchboard corpus to determine a \textit{human word error rate} and also to uncover correlations between transcription errors by humans and ASR systems on this data.

Inspired by these works, in the following experiment, we estimate a \textit{human word error rate} on transcribing German oral history interviews. Strictly speaking, this human error rate is the difference of transcriptions between two transcribers. One transcriber is taken as the reference and the other as the hypothesis. These results aim to expose the challenges even humans face transcribing oral history interviews and put the achieved error rates of speech recognition systems in this domain into perspective. 

\newcite{Xiong.2017.HumanPartiyInASR} used a two-staged transcription pipeline by a large commercial vendor to transcribe the English Switchboard data. In the first stage, the pre-segmented speech is annotated by one professional transcriber from scratch. In the second stage, a second transcriber corrects the first transcriber's transcription. Using this approach, the authors report a 5.9\,\% human error on the English Switchboard data.

Later on, \newcite{Saon.2017.IBM_SWBD} replicated the experiment on the same data as the authors consider the values reported by \newcite{Xiong.2017.HumanPartiyInASR} to be too high. The authors also used a two-staged pipeline by a large commercial vendor, where three different transcribers transcribed the speech segments from scratch independently from each other. Then, in the second stage, a fourth transcriber performed a quality check and corrected the annotations of the first stage. \newcite{Saon.2017.IBM_SWBD} report a human word error rate in the range of 5.1--5.6\,\% on English Switchboard.

With the project's budget, we could not apply such a professional commercial pipeline. Instead, we aim at approximating the two-stage procedure by replacing the first stage with our speech recognition system. For prepossessing the HdG data set for annotation, we apply our current automatic transcription system (Fraunhofer IAIS Audio Mining) with our latest robust broadcast ASR model to create a raw ASR transcript, including punctuation. \newcite{Schmidt.2016.AudiominingCurrent} give an overview of the Audio Mining system and the audio analysis pipeline. The ASR model is a slightly improved version of the model proposed in \cite{Gref.2019.twostaged_am_adaption}. For the LF-MMI acoustic model \cite{Povey.2016.LF_MMI}, 1,345 hours of training data from the broadcast domain were used and 9-fold data augmented (3-fold speed perturbation and 3-fold noise and reverberation data augmentation). In a second stage, the model was fine-tuned to oral history interviews with 3.5 hours of data. For the language model, broadcast data was combined with manual oral history transcripts in roughly equal proportions.

We use the ASR result to chunk the interviews into short segments at the longest speech pauses until we obtain segments of 30 seconds or less. We obtain 3,122 segments for the 10 hours of data by this approach. Thus, the average segment length for our data is 11.5 seconds. 

In the second stage, three human transcribers were provided with the same raw ASR transcript of the 10 hours of speech and were asked to correct it independently. This was performed by three employees at the \textit{Haus der Geschichte} who have an academic background in history---but are not professional transcribers. The transcribers did not only correct the ASR transcription but also annotated the perceived emotions and sentiment for each segment. We use these annotations in an additional future study of emotions and sentiment on oral history.

\subsection{Results for Human Word Error Rate}

\begin{table}[t]
	\centering
	\begin{tabular}{cclrrr}
		\toprule
		&				&							&   \multicolumn{3}{c}{\textbf{Hypothesis}} \\ \cmidrule(l{5pt}r{5pt}){4-6}
		& \textbf{WER Setup} 	&	\textbf{Ref.}		&	\textbf{Tr. A}	& \textbf{Tr. B}	& \textbf{Tr. C} \\
		\midrule
		&	\multirow{3}{*}{\shortstack{CS\\[4pt]Hesitations\\counted}}		
		& Tr. A 				&	-		& 7.77	& 9.51		\\
		1)	&	& Tr. B 		&	7.79	& -			& 8.83	\\
		&	& Tr. C 			&	9.54	& 8.83	& -		\\		
		\midrule
		&	\multirow{3}{*}{\shortstack{CS\\[4pt]Hesitations\\ignored}}	
		&  Tr. A 				&	-		& 7.41	& 9.00		\\
		2)	&	& Tr. B 		&	7.45	& -			& 8.58	\\
		&	& Tr. C 			& 	9.06	& 8.58	& -		\\		
		\midrule
		&	\multirow{3}{*} {\shortstack{CI\\[4pt]Hesitations\\ignored}}	
		& Tr. A 				&	-		& 6.54	& 8.43		\\
		3)	&	& Tr. B 		&	6.58	& -			& 7.74	\\
		&	& Tr. C 			& 	8.48	& 7.75	& -		\\		
		\bottomrule
	\end{tabular}
	\caption{Overview of the human word error rates in percent between two different transcribers (Tr.): A, B, and C. Three different types of experiments were performed to investigate different reasons for the resulting error rates. CS: case-sensitive, CI: case-insensitive.}
	\label{tab:hdg_human_wer_overview_big_table}
\end{table} 

The results of comparing the different human transcriptions are summarized in Table \ref{tab:hdg_human_wer_overview_big_table}. We compared three different setups, one after the other, to investigate different causes for the error rates. 

We begin with Setup 1 in the top third of Table \ref{tab:hdg_human_wer_overview_big_table} that we usually also consider when evaluating speech recognition systems: The word error rate is calculated case-sensitively, i.e., different casings of the same words are counted as substitutions in the WER. Furthermore, transcribers were asked to transcribe hesitation sounds with a predefined spelling. Our ASR system usually transcribes these hesitation sounds if they can be heard clearly enough. The highest difference is between transcriber A and C, with a 9.5\,\% word error rate with this setup. The lowest is between transcriber A and B, with a 7.8\,\% word error rate.

For a more detailed analysis, we consider the combination with the lowest error rate in this setup: Transcriber A as the reference and B as the hypothesis. 
%Transcriber A has overall 78,428 transcribed words that are used as the reference. Comparing Transcriber B to A, Transcriber B has 6093 errors---1106 insertions, 1328 deletions, and 3659 substitutions. 
An overview of the top five errors for each category is given in Table \ref{tab:hdg_human_wer_error_ops_stats}. The most common errors are insertions of the hesitation sounds \textit{äh} (German variant of the hesitation \textit{er} or \textit{uh}), \textit{Äh}, and \textit{hm}, by Transcriber B. It seems B was paying more attention to hesitation sounds than A. However, it is noteworthy that both transcribers have the same annotation of hesitation sounds way more often than not. Both transcribed the most common hesitation sound \textit{äh} at the same position in 675 cases. Overall in 159 cases, \textit{äh} led to an error.

\begin{table}[t]
	\centering
	\begin{tabular}{lrrr}
		\toprule
		\textbf{Error} 	&	\textbf{Tr. A} 		& \textbf{Tr. B}	 & \textbf{Error}  \\
		\textbf{Type}	& 	\textbf{(ref)}		& \textbf{(hyp)}	& \textbf{Count} \\
		\midrule
		Deletion        &           und         &        -     & 71 \\
		Deletion        &            ja         &        -     & 63 \\
		Deletion        &           ich         &        -     & 47 \\
		Deletion        &          dann         &        -     & 37 \\
		Deletion        &            in         &        -     & 37 \\
		\midrule
		Insertion       &           -         &         äh    & 118 \\
		Insertion       &           -         &         hm    &  50 \\
		Insertion       &           -         &         Äh    &  38 \\
		Insertion       &           -         &        und    &  35 \\
		Insertion       &           -         &        die    &  25 \\
		\midrule
		Substitution    &          habe         &        hab    &  61 \\
		Substitution    &           sie         &        Sie    &  61 \\
		Substitution    &          dass         &        das    &  43 \\
		Substitution    &           ich         &        Ich    &  39 \\
		Substitution    &           und         &        Und    &  39 \\	
		\bottomrule
	\end{tabular}
	\caption{Top five errors for each error type between Transcriber A and B for a case-sensitive human word error rate estimation on German oral history data where hesitation sounds are annotated and counted as word errors.}
	\label{tab:hdg_human_wer_error_ops_stats}
\end{table}

The next common error type comparing A and B is deletions of short words---\textit{und} (\textit{and}) and \textit{ja} (\textit{yes})---that A has annotated quite often, but B has not. The next most common errors are substitutions of the same words in slightly different spellings: e.g., formal \textit{habe} vs.\ informal \textit{hab} (\textit{have}), and casing errors.

These observations lead us to two questions that will be investigated with two further setups: what is the influence of hesitation sounds on the error rate? And what is the influence of the casing? 

To answer the first question, we remove the hesitation sounds from the transcript of all transcribers. The resulting error rates are depicted as Setup 2 in the middle part of Table \ref{tab:hdg_human_wer_overview_big_table}. Overall, without hesitation sounds, the human WER is decreased by 0.3--0.5 percentage points. Since the sounds were removed from both the reference and hypothesis, the overall influence on the human error rate is quite limited.

\begin{table}[t]
	\centering
	\begin{tabular}{lrrr}
		\toprule
		& \textbf{Setup 1}	& \textbf{Setup 2}	& \textbf{Setup 3} \\
		\midrule
		Casing		& counted	& counted	& ignored \\
		Hesitations	& counted	& ignored	& ignored \\
		\midrule
		WER			& 7.77\,\%	& 7.41\,\%	& 6.54\,\% \\
		Num.\ Ref.	& 78,428	& 77,383	& 77,383 \\
		Error Sum	& 6,093		& 5,733		& 5,059 \\
		Insertions	& 1,106		&   835		&   835 \\
		Deletions		& 1,328		& 1,288		& 1,288 \\
		Substitutions		& 3,659		& 3,610		& 2,936 \\
		\bottomrule
	\end{tabular}
	\caption{Word error count comparison between Transcriber A (ref) and Transcriber B (hyp) for the three different human error rate investigation setups.}
	\label{tab:hdg_human_wer_stats_A_vs_B}
\end{table} 

\begin{table}[t]
	\centering
	\begin{tabular}{lrrr}
		\toprule
		\textbf{Error} 	&	\textbf{Tr. A} 		& \textbf{Tr. B}	 & \textbf{Error}  \\
		\textbf{Type}	& 	\textbf{(ref)}		& \textbf{(hyp)}	& \textbf{Count} \\
		\midrule
		deletion	& und	& -	& 73 \\
		deletion	& ja	& -	& 64 \\
		deletion	& ich	& -	& 47 \\
		deletion	& dann	& -	& 39 \\
		deletion	& in	& -	& 38 \\
		\midrule
		insertion	& -	& und	& 35 \\
		insertion	& -	& die	& 26 \\
		insertion	& -	& da	& 24 \\
		insertion	& -	& ich	& 22 \\
		insertion	& -	& dann	& 20 \\
		\midrule
		substitution	& habe	& hab	& 62 \\
		substitution	& dass	& das	& 43 \\
		substitution	& das	& dass	& 35 \\
		substitution	& dann	& da	& 18 \\
		substitution	& das	& es	& 18 \\		
		\bottomrule
	\end{tabular}
	\caption{Top five errors for each error type between Transcriber A and B for a case-insensitive human word error rate estimation with ignored hesitations on German oral history data.}
	\label{tab:hdg_human_wer_error_ops_stats_clean}
\end{table} 

Comparing A with B again, the total number of errors reduces by 360 errors by removing hesitation sounds, as reported in Table \ref{tab:hdg_human_wer_stats_A_vs_B}. The largest share of this is accounted for by an improvement in insertions---as may already be assumed in advance. However, at the same time, the total number of words in the reference decrease even more by more than 1000 words. 

Lastly, we examine the influence of casing errors comparing the transcriptions with removed hesitations and lower-casing. The results are depicted as Setup 3 in the bottom third of Table \ref{tab:hdg_human_wer_overview_big_table}. Ignoring the casing for evaluation reduces the word error rate by a further 0.5 to just under 1.0 percentage points. As shown in Table \ref{tab:hdg_human_wer_stats_A_vs_B}, ignoring the casing naturally reduces only the number of substitutions.

As shown in Table \ref{tab:hdg_human_wer_error_ops_stats_clean}, after removing the hesitation transcriptions and lower-casing all words, the top five inserted words by Transcriber B are now also mostly short words with only one syllable---words that can be easily overheard in spontaneous speech, especially when there are word repetitions or ungrammatical sentences due to rephrasing. However, the top five errors per category account for just under 11\,\% of all word errors with this setup. Therefore, a large share of the errors is distributed among many individual errors that are not as systematic as these.

Finally, we take the arithmetic mean of the six different transcriber pairs to report a human word error rate for each of the three different analysis scenarios we studied. These values are given in Table \ref{tab:hdg_human_wer_error_overview}, in addition with the standard deviation. 

For the evaluation we perform in our research---case-sensitive word error rate evaluation and annotating hesitations---the corresponding human word error rate on oral history interviews is 8.7\,\%. There are two primary reasons why we evaluate our ASR with Setup 1. For the indexing of the content and adequate readability, the casing of words in the German language is crucial and should be correctly transcribed by systems. In our ASR system, the casing is part of the language model and pronunciation lexicon since we achieve better overall recognition results than with a downstream inverse-text-normalization component. Thus, we evaluate the casing as a substitution in our ASR. Additionally, as mentioned at the beginning of this section, transcribed hesitations are crucial in specific research questions. They help assess not only what but how something was said. Therefore, we also evaluate these hesitations in our ASR evaluation.

\begin{table}[t]
	\centering
	\begin{tabular}{lr}
		\toprule
		\textbf{Variant of Analysis}		& \textbf{Avg.\ Human WER} \\
		\midrule
		\begin{tabular}{l}
			Case-sensitive WER,\\
			including hesitations \\[4pt]
		\end{tabular} & 
		\begin{tabular}{l}
			8.71\,\% $ \pm $ 0.79\,\% 
		\end{tabular} \\
		\begin{tabular}{l}
			Case-sensitive WER,\\
			excluding hesitations \\[4pt]
		\end{tabular} & 
		\begin{tabular}{l}
			8.35\,\% $ \pm $ 0.74\,\% 
		\end{tabular} \\[4pt]
		\begin{tabular}{l}
			Case-insensitive WER,\\
			excluding hesitations \\
		\end{tabular} & 
		\begin{tabular}{l} 
			7.59\,\% $ \pm $ 0.86\,\% 
		\end{tabular} \\
		\bottomrule
	\end{tabular}
	\caption{Average human word error rates on 10 hours of manually transcribed German oral history interviews. The average human word error rate is given as the arithmetic mean $ \pm $ the standard deviation of the six different comparisons in each setup.}
	\label{tab:hdg_human_wer_error_overview}
\end{table} 

\begin{table*}[t]
	\centering
	\begin{tabular}{llllll}
		\toprule
				\textbf{}		&  & \multicolumn{4}{c}{\textbf{Adaptation Data}}  \\ 
			\cmidrule(l{5pt}r{5pt}){3-6} 	
		\textbf{	}		& \textbf{Broadcast} 	&  \textbf{HdG Train} &   \textbf{}  		& \textbf{HdG Train A}	&	 		\\ 
		\textbf{Set}		& \textbf{Baseline} 	&  \textbf{Transc.\ A} 	  &   \textbf{KA3 25h*} & \textbf{+ KA3 25h*}	& \textbf{KA3 250h*}	\\ 
		\midrule	
		KA3 Test			& 26.0	& 25.7	& 24.7			& 24.6			& \bftab{23.9} 	\\
		\midrule 
		HdG Dev.\ Avg.		& 17.3 $ \pm $	1.06 & 17.0$ \pm $	1.08	& 16.7$ \pm $	1.02	& \bftab{16.6}$ \pm $ 1.08	& 17.1 $ \pm $	1.09	\\
		~ Transcriber A 	& 16.7	& 16.4	& 16.2			& \bftab{16.0}	& 16.4 	\\
		~ Transcriber B		& 16.6	& 16.3	& 16.1			& \bftab{16.0}	& 16.5 	\\
		~ Transcriber C		& 18.5	& 18.2	& \bftab{17.9}	& \bftab{17.9}	& 18.4 	\\
		\midrule
		HdG Test Avg.		& 16.4 $ \pm $	0.32 & 15.9	$ \pm $	0.30 & \bftab{15.6}	$ \pm $ 0.33 & 15.7	$ \pm $	0.37 & 16.1 $ \pm $	0.36 \\
		~ Transcriber A		& 16.1	& 15.6	& \bftab{15.3}	& \bftab{15.3}	& 15.8 	\\
		~ Transcriber B		& 16.4	& 15.9	& \bftab{15.6}	& 15.7			& 16.1 	\\
		~ Transcriber C		& 16.8	& 16.2	& \bftab{16.0}	& 16.1			& 16.5 	\\
		\bottomrule
	\end{tabular}
	\caption{Comparison of acoustic model adaptation experiments using different oral history adaptation data sets. Results are reported as word error rate in percentage. HdG Test and Dev.\ Average (Avg.) are the respective arithmetic mean $ \pm $ the standard deviation of the results of the ASR system on the three different human transcriptions. Additionally, we also report the respective results for Transcriber A, B, and C as the reference. \textit{Broadcast} is our recent base model. The KA3 adaptation data marked with an asterisk (*) are not manually annotated. It was created using forced alignment of raw human transcription, automatic segmentation and data clean-up.}
	\label{tab:asr_wer_results}
\end{table*} 

\subsection{Discussion and Limitations}

Compared to the 4.0---5.9\,\% human word error reported for English Switchboard, the human error rates on German oral history data we obtained are significantly higher. This is especially because of the characteristics of oral history interviews, which were pointed out in the related work section.

At the same time, we must admit that the experiment is subject to some limitations. The calculated error rate depends on the transcribers, their motivation, and the applied procedure. On the one hand, presumably, it would be possible to reduce the differences between the transcribers by several correction iterations. On the other hand, we find it remarkable that all transcribers had the same raw ASR transcript as a basis---and yet such comparatively large differences in the annotations can be found. It can be assumed that applying an annotation from scratch would result in an even higher error rate. 

Another advantage that humans have over the ASR system is the context. The transcribers were aware of the content discussed in the interviews and could listen to previous and subsequent segments at will. Our ASR system does not have this advantage and transcribes segment by segment independently. The ASR system would need to account for surrounding segments---which is currently not supported for standard n-gram language models---for a fair comparison. 

If the human transcribers had listened and annotated each segment in a random order, this would naturally result in a higher human word error rate. Finally, the transcribers were provided not only with the audio but also with the video stream. It is well-known that visual feedback, e.g., seeing the lip movement, can improve speech understanding. This can be another advantage for human transcription. Nevertheless, based on the annotator's feedback, transcribing only audio segments (especially if in random order) would have resulted in significantly reduced motivation, which in turn would have spuriously affected transcription quality.

Lastly, it should be emphasized that the oral history interviews used for this experiment have a fairly high audio quality and are quite easy to understand. This is not true to the same extent for many other oral history interview archives in different languages. For instance, the KA3 interviews we examined in our prior works have much more challenging acoustic conditions. For these interviews, a significantly higher human error rate can be assumed. We also observe this effect with ASR systems, which, despite robust training, often struggle with interviews with poor audio quality.

\section{ASR Results}

In this section, we evaluate our recent ASR model on the HdG data set transcribed in the previous section and perform adaptation experiments based on \cite{Gref.2019.twostaged_am_adaption} with different oral history training data sets. We use the same adaptation setup for all data sets, particularly the same fixed number of four epochs and the same initial learning rate and decay, as given in the paper. The same neural network architecture and training routine as in the previously mentioned work is used for the experiments: an LF-MMI trained acoustic model \cite{Povey.2016.LF_MMI} with three LSTM and seven TDNN layers.

Our recent broadcast acoustic model serves as the baseline for the experiments. The model is trained on 1,345 hours speech from the broadcast domain that is 9-fold data augmented (3-fold speed perturbation and 3-fold noise and reverberation data augmentation). This model achieves near-human performance on broadcast speech: 8.8\,\% WER on planned speech in clean acoustic conditions and 9.8\,\% in noisy, both with non-restricted vocabulary and a general-purpose broadcast language model. For academic purposes, this baseline ASR model can be used for free with a limited monthly contingent as part of the \textit{BAS Speech Science Web  Services} \cite{Kisler.2016.BASWebservice}.\footnote{\url{https://clarin.phonetik.uni-muenchen.de/BASWebServices}}.

For adaptation, we use the 6.35-hour training subset of the HdG presented in this paper. Additionally, we use data sets created with an approach used in our prior works \cite{Gref.2020.MultiStageCrossLingual} in the KA3 project, where we applied forced alignment and automatic segmentation with data clean-up on human transcriptions of oral history interviews. On average, the KA3 interviews used for this task are much older recordings than the HdG data and, therefore, often have poorer recording quality. We use a data set with 250 hours of aligned interviews from 150 different speakers and a 10\,\% subset with 25 hours. With the 25-hour subset, we investigate the effect of a possible domain-overfitting with 250 hours on the acoustic conditions of KA3. We further investigate the combination of this \textit{noisy} data set with the comparability \textit{clean} HdG training data set. For evaluation, we use a large 5-gram general-purpose broadcast language model with over 2 million words in the lexicon. The overall results are given in Table \ref{tab:asr_wer_results}.

The experiments show that adaptation of the acoustic model can lead to consistent ASR improvements. However, the overall improvement is dependent on the data, and more data does not necessarily lead to better recognition performance. With the 250 hours of forced aligned data for adaptation, we achieve a relative improvement of 8.2\,\% on KA3 Test, compared to the broadcast base model. This model outperforms our prior best model \cite{Gref.2020.MultiStageCrossLingual} by 1.3 percentage points. However, on the HdG development and test set, the relative improvement of this model on HdG Dev.\ (1.1\,\%) and Test (2.0\,\%) is quite consistent but comparatively small.  

In contrast, adaptation on the training split of Transcriber A results in 2.0\,\% improvement on HdG Dev and 3.2\,\% on Test---although it is only 6.35 hours of training data and thus magnitudes smaller than the 250 hours KA3 set. The adaptation improves not only using A as the reference for the WER, but also very similar for Transcriber B and C. Thus, the improvement on the HdG data is consistent and not just a bias towards the transcription style of A. However, the improvement of this adaptation on KA3 test is only 1.1\,\% relative to the baseline.

An even better result on HdG compared to training with the small HdG set is obtained with the 25-hour KA3 dataset. Although this dataset has much worse acoustic recording conditions and contains semi-automatic transcriptions, the acoustic model seems to generalize better with this data. For HdG Dev, the relative improvement here is 3.2\,\% relative to baseline, and for HdG Test 4.9 \%. On KA3 Test, we achieve a relative improvement in a similar range of 5.1\,\%. Combining the HdG with the 25-hour KA3 training data set slightly improves further the performance on the KA3 Test, and the HdG Dev set. However, on HdG Test, the performance is somewhat decreased. Again, this is consistent for all reference transcribers. 

Overall, we can conclude that substantial improvement on oral history data can be achieved with comparatively few hours of adaptation data, both semi-automatically and manually annotated. A large data set can lead to overfitting to the domain, as in the case of KA3. Thus, depending on the application and the type of data, it may be valuable to experiment with varying subsets of data. Adjusting the learning rate for large subsets is also a possibility that could be explored. However, this is often at the expense of training time. Overall, a relative improvement of about 5\,\% can be achieved by adapting the acoustic model for each of the two different oral history test data. 

Compared to the human word error rate of 8.7 percent that we worked out in this paper, ASR on the HdG data set still has quite a way to go to achieve human performance on oral history data. The error rate has to be roughly halved until an ASR system can replace manual transcription in most scenarios and make human transcriptions superfluous. However, the current recognition performance of the systems is already sufficient so that after a manual correction, the transcript can be used for the \textit{Zeitzeugenportal} of the \textit{Haus der Geschichte} Foundation. The transcripts are essential documents for the practical use of oral history interviews. They are primarily used for indexing the contents of the videos for the thematic classification on the online service \textit{Zeitzeugenportal}. Since oral history videos are also a meaningful component of the exhibition practice in museums, the transcripts are also used for cut lists. Additionally, the transcripts serve for the subtitling of the videos.

Furthermore, the oral history interviews from the two different archives, HdG and KA, uncover a substantial difference in speech recognition performance. The KA3 test data is much more challenging for the ASR system, even when adapted with 250 hours of additional representative data from the very same archive. Depending on the model, the absolute difference in WER between KA3 Test and HdG Test is 9.5\,\% (Broadcast Baseline), 7.7\,\% (KA3 250 hour), and 8.9--9.0\,\% (KA3 25 hour with/without HdG train). Both data sets are from the German oral history domain have similar characteristics of speakers, especially in terms of age, language, dialects, and topics. The main difference lies in the wide range of recording age of the KA3 interviews and the resulting acoustic recording quality. Therefore, for our models, approximately 9\,\% of WER percentage points are still attributable to the acoustic challenges of oral history. Thus, further improving acoustic robustness for oral history remains an open field of research, although there has been substantial improvement in recent years.

\section{Conclusions}

In this work, we investigated the accuracy of human transcription of German oral history data by comparing corrected versions of raw ASR transcription of three different persons. We estimate a word error rate of 8.7\,\% for recent oral history interviews with relativity clean acoustic conditions. We discussed the different types of possible human transcription errors of oral history interviews. This error rate is intended to serve us as a rough benchmark for estimating human transcription accuracy and is by no means to be taken as an absolute benchmark of human performance. We have discussed the limitations of our approach and argued that different approaches to estimating the human error rate can lead to different results---as has been the case, for example, with several works on the human error rate estimation on the Switchboard corpus in the past. We suspect our human error rate for oral history is more likely to be on the low end and may be much higher on the same data when transcribing from scratch or using a random order of segments. However, we think that our error rate estimate can serve as a rough reference when assessing ASR systems on oral history---and what realistic word error rates of ASR systems can be expected in the future.

In this paper, we have further presented experiments on the adaptation of acoustic models for oral history with different data sets. We discussed the influence of the training data set for adaptation. We showed that a large amount of data might not necessarily lead to good generalization but instead might lead to domain overfitting. Even with just 25 hours of data, a consistent improvement by 5\,\% relative to a robust baseline can be achieved for oral history.

\section{Bibliographical References}\label{reference}
%\label{main:ref}

\bibliographystyle{lrec2022-bib}
\bibliography{paper}

\section{Language Resource References}
\label{lr:ref}
\bibliographystylelanguageresource{lrec2022-bib}
\bibliographylanguageresource{languageresource}

\end{document}